\documentclass[11pt]{article}

\usepackage{array}
\usepackage{subcaption}
\usepackage{subdepth}
\usepackage{url}
\usepackage{fullpage} 
\usepackage{graphicx}
\usepackage{authblk}
\usepackage{amsmath}

\renewcommand{\arraystretch}{1.25}
\newcommand{\bhline}{\noalign{\hrule height 0.8pt}}

\title{Compressing Sign Information in DCT-based Image Coding via Deep Sign Retrieval}

\author[1]{Kei Suzuki}
\author[1]{Chihiro Tsutake}
\author[1]{Keita Takahashi}
\author[1]{Toshiaki Fujii}

\affil[1]{
Department of Information and Communication Engineering, 
Nagoya University, 
Furo-cho, 
Chikusa-ku, 
Nagoya, 
464-8603, 
Japan}

\date{\empty}

\begin{document}

\maketitle
\vspace{-10mm}

\begin{abstract} 
Compressing the sign information of discrete cosine transform~(DCT) coefficients is an intractable problem in image coding schemes due to the equiprobable characteristics of the signs. To overcome this difficulty, we propose an efficient compression method for the sign information called ``sign retrieval.'' This method is inspired by phase retrieval, which is a classical signal restoration problem of finding the phase information of discrete Fourier transform coefficients from their magnitudes. The sign information of all DCT coefficients is excluded from a bitstream at the encoder and is complemented at the decoder through our sign retrieval method. We show through experiments that our method outperforms previous ones in terms of the bit amount for the signs and computation cost. Our method, implemented in Python language, is available from \url{https://github.com/ctsutake/dsr}.
\end{abstract}


\section{Introduction}
The discrete cosine transformation~(DCT)~\cite{Ahmed1974discrete} is known as an important technique for image coding and is adopted in various image coding standards~\cite{Wallace1991jpeg,Pater2005Berkeley,Puri1993video,Rijkse1996h263,Hannuksela2015high,Wiegand2003overview,Sullivan2012overview,Bross2021overview}. For instance, JPEG~\cite{Wallace1991jpeg} first divides an original image into non-overlapping blocks and then applies DCT to each of the blocks followed by quantization. Entropy coding is finally performed to obtain bit representations for the quantized DCT coefficients. According to the source coding theory~\cite{Shannon1948math}, statistically biased symbols can be efficiently compressed using entropy coding methods such as \cite{Huffman1952method,Golomb1966run,Rice1971adaptive,Martin1979range}. However, the sign information of DCT coefficients has equiprobable characteristics~\cite{Reininger1983distribution,Bellifemine1992statistical,Lam2000math}, i.e., the probabilities of the positive and negative signs are almost even, and the compression of the sign information has been thus considered impossible. Therefore, each of the signs is represented using $1$ bit in typical image coding methods; the sign information consumes many bits in the resulting bitstream. To reduce the bit amount for the signs, we address a sign compression problem for DCT coefficients in this paper. In particular, we consider a lossless sign compression problem, where the signs of the DCT coefficients are decoded without loss.

\begin{figure}[!t]
\centering
\includegraphics[width=120mm]{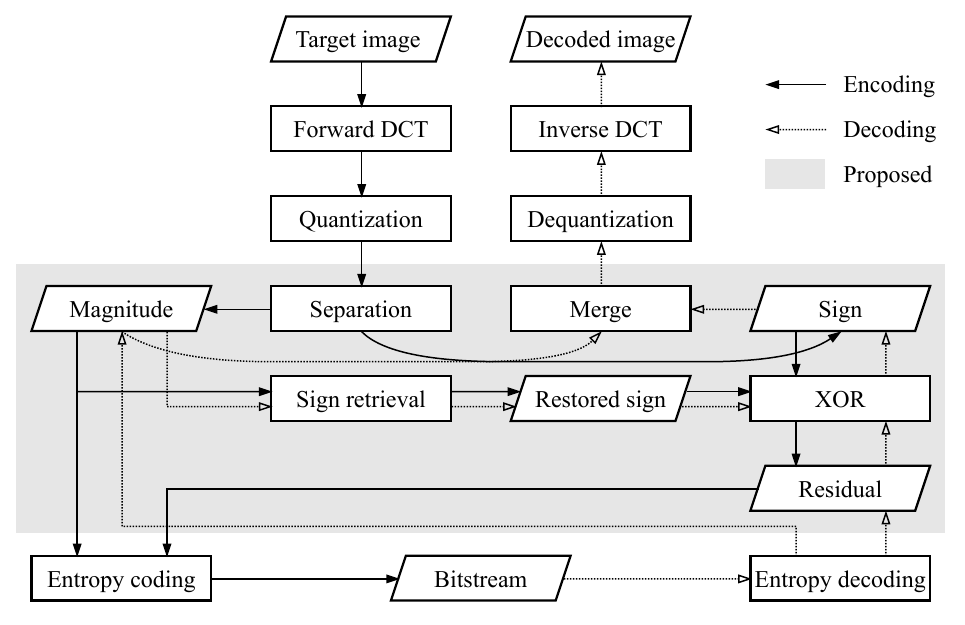}
\caption{Our encoder and decoder.}
\vspace{4mm}
\label{fig:encdec}
\end{figure}

\begin{figure}[!t]
\centering
\includegraphics[width=120mm]{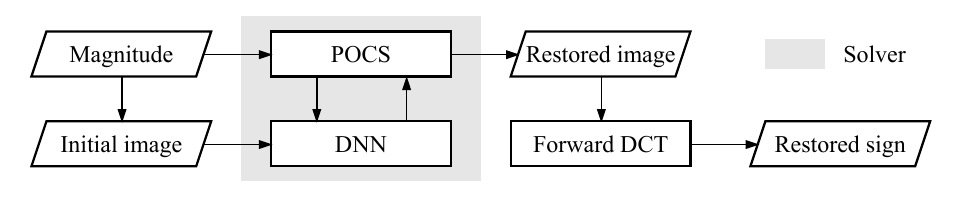}
\caption{Sign retrieval.}
\label{fig:sr}
\end{figure}

We briefly summarize seminal works developed to tackle this challenging problem. The previous methods can be classified into prediction-based~\cite{Ponomarenko07prediction,Kazui2010description,Lakhani2013modifying,Henry2016residual,Nakagawa2017sign,Filippov2019residual} and context-based \cite{Tu2002context,Miroshnichenko18compression} methods. The prediction-based methods consist of several steps. The pixel values of a target block are first predicted from those of the already encoded and decoded neighboring blocks. The predicted result is further used to predict the signs of the DCT coefficients for the target block. Since some of the predicted signs might be incorrect, the residual information for correcting the errors is also computed and appended to the bitstream. On the other hand, the context-based methods exploit the fact that the signs are correlated with each other in the same and neighboring blocks. A conditional probability model for the signs is constructed and incorporated into the entropy coding.

To more efficiently compress the sign information, we propose a novel method significantly different from the current literature. Our method is inspired by a classical signal processing method, referred to as phase retrieval~\cite{Gerchberg1972practical,Fienup1982phase,Fickus14phase,Candes2015phaseA,Candes2015phaseB,Goldstein2018phasemax,Bahmani2017flexible,Salehi18learning}, which tries to find the phase information of discrete Fourier transform~(DFT) coefficients only from their magnitudes. Our method is called ``sign retrieval'' because the sign information of the DCT coefficients is restored only from their magnitudes. Sign retrieval is incorporated into a typical pipeline for image coding such as JPEG as shown in Fig.~\ref{fig:encdec}; sign retrieval is included in the gray region, and it is combined with other components, i.e., DCT, quantization, and entropy coding, which are commonly used for image coding. To encode a target image, the encoder separates the quantized DCT coefficients into the magnitude and sign information. We then restore the signs only from the magnitudes via sign retrieval. We also compute the residual information, that is, the difference between the true signs of the DCT coefficients and the restored ones. We finally compose a bitstream that includes the magnitudes and the residual. The decoder performs the reverse operations to obtain the decoded image. 

Figure~\ref{fig:sr} summarizes sign retrieval. It is formulated as an optimization problem with specific constraints in the DCT domain, and it is solved using deep neural networks~(DNNs)~\cite{Liu2017survey}. We first compute an initial image from the magnitude information as the input for our solver.  The solver, which corresponds to the gray region in Fig.~\ref{fig:sr}, consists of two building blocks; the first is a DNN whose input and output are images, and the second is the projection onto convex set~(POCS) operator~\cite{Bauschke1996projection} for enforcing the constraints imposed by the DCT magnitudes.  Taking the initial image as the input, the solver alternates between DNN inference and POCS, similar to iterative methods in numerical optimization~\cite{Nocedal2006numerical}. The initial image is gradually refined while this alternation is repeated many times. It is expected that, after a sufficient number of iterations, a restored image can be obtained that has mostly correct signs in the DCT domain. We finally obtain the restored sign information from the restored image. 

A preliminary version of our method was presented in conference proceedings~\cite{Tsutake21efficient}, and we here enhance the algorithm by using a DNN as a building block. We show through experiments that our method outperforms not only our preliminary method but also other previous methods~\cite{Nakagawa2017sign,Tu2002context,Miroshnichenko18compression} in terms of the bit amount for the sign information and computation cost. 

The rest of this paper is organized as follows. In Section~\ref{s2}, we briefly review related work on sign compression. In Section~\ref{s3}, we elaborate sign retrieval and its application to image coding. In Section~\ref{s4}, we report the experimental results including comparisons with other methods. We conclude in Section~\ref{s5} with a brief summary and mention of future work.

\section{Related work}
\label{s2}
\subsection{Prediction-based sign compression}
\label{s2ss1}
In a series of papers, Ponomarenko et al.~\cite{Ponomarenko07prediction} pioneered the framework of prediction-based sign compression methods. The magnitude information was assumed to be encoded and decoded before sign compression. To compress signs, they first estimate the pixel values of a target block using adaptive filtering~\cite{Golchin1997context}. More specifically, they compute the weighted sum of already encoded, decoded, and interpolated pixel values in raster-scan order. The signs in the target block are then predicted coefficient-by-coefficient by minimizing a cost function with respect to the signs. The prediction error, i.e., the residual information, is finally transmitted to correct the signs at the decoder side. It was demonstrated that the bit amount of the residual is only $60$--$85$\%  of the original sign information. To reduce the computation cost for sign prediction, the authors of \cite{Henry2016residual,Nakagawa2017sign} introduced look-up table methods, and Filippov et al.~\cite{Filippov2019residual} removed inverse DCT from the cost function.

In the Joint Video Experts Team (JVET), similar kinds of prediction-based methods have gained significant attention because of their good compression performance. Several variants~\cite{Auyeung2021low,Xu2022adaptive,Xiu2022enhanced} have been actively proposed for a new video compression standard beyond versatile video coding (VVC)~\cite{Bross2021overview}. In \cite{Chen2022more,Naser2022code}, sign compression methods were developed for transformations other than DCT such as discrete sine transformation~\cite{Jain1976fast} and low frequency non-separable transformation~\cite{Koo2019low}.

\subsection{Context-based sign compression}
\label{s2ss2}
Tu and Tran~\cite{Tu2002context} proposed a context-based method inspired by sub-band coding~\cite{Shapiro1993embedded,Said1996new,Taubman2000high}. They compose blocks, each of which contains DCT coefficients belonging to the same frequency band. They then compress the sign information using context-based adaptive binary arithmetic coding~\cite{Marpe2003context}. As much as $36$ context models are switched in accordance with the signs in the same and neighboring blocks. The compression efficiency is slightly better than the case without their context modeling. Miroshnichenko et al.~\cite{Miroshnichenko18compression} proposed another method, where not only the signs but also the magnitudes are used to estimate the conditional probability among them. The bit amount is about $80$\% that of the case without the conditional probability. Although context-based methods require no inverse DCT operations, which is advantageous in terms of computation cost, the compression efficiency is limited compared with prediction-based methods.

\subsection{Prediction-based joint sign and magnitude compression}
\label{s2ss3}
References~\cite{Puri1997improvements,Lakhani2007dct,Lakhani2008image,Swamy2008adaptive} aimed to improve the compression performance for DCT coefficients by predicting not only the signs but also the magnitudes. In other words, these methods predict DCT coefficients themselves. The origin of these methods dates back to \cite{Puri1997improvements}, whose details are given below. The DC component of a target block is predicted from those in either the left or above neighboring blocks. Depending on an empirical criterion, the AC components are also predicted in a similar manner. The prediction residual is transmitted as was done in \cite{Ponomarenko07prediction,Kazui2010description,Lakhani2013modifying,Henry2016residual,Nakagawa2017sign,Filippov2019residual}. This method was applied to MPEG-1~\cite{Pater2005Berkeley}, MPEG-2~\cite{Puri1993video}, and H.263~\cite{Rijkse1996h263} video coding methods, and the bit amount was reduced to $77$\% of the baseline for H.263. Swamy and Kumar~\cite{Swamy2008adaptive} first encode and decode DC components and then predict the AC components as the weighted sums of the DC components. The weights are learned from a set of training images using linear programming~\cite{Danzig1963linear}. According to \cite{Swamy2008adaptive}, prediction of DCT coefficients can be regarded as surface reconstruction, where only smooth surfaces can be obtained. Therefore, high-frequency components, which are not smooth, are difficult to predict with certain kinds of the method in \cite{Swamy2008adaptive}.

\subsection{Sign data hiding}
\label{s2ss4}
Sign data hiding, originally proposed by Clare et al.~\cite{Clare2011sign}, is a building block of high efficiency video coding (HEVC)~\cite{Sullivan2012overview} and is a fundamentally different approach from those reviewed so far. Sign data hiding skips (does not encode) the sign of the first significant AC coefficient in a target block. At the decoder side, the missing sign is determined depending on a parity check (the sum of all the magnitudes). To ensure that a sign determined in this way should always be correct, the magnitudes are slightly modified at the encoder side. This approach is extended in a series of papers~\cite{Wang2012multiple,Zhang2013additional,Song2018extra}, where more than a single sign per block can be skipped. However, the original magnitudes are modified for the parity check, resulting in degraded image quality.

\subsection{Image coding and deep neural network}
Recent years have seen growing interest in applying DNN-based methods to image coding. These methods can be classified into two groups. The methods in the first group~\cite{Balle2017end,Balle2018variational,Cheng2020learned,Cui2021asymmetric,Li2021advances} model the entire image coding pipeline as a DNN and optimize it end-to-end on a set of training images. The bitstream is generated from the latent features that are extracted from a target image using the DNN. These methods achieve remarkable performance in the low-bitrate range, but the performance is limited for the high-bitrate range compared with traditional image coding methods. The methods in the second group~\cite{Pfaff2018neural,Huo2018convolutional,Schafer2019affine,Dumas2020iterative} use DNNs as modules in the traditional image coding pipeline. Some of the building blocks, such as those for spatial and temporal predictions, are replaced or enhanced by DNNs trained on a set of training images. These methods are successful throughout the bit-rate range, yielding consistently better rate-distortion performance than their counterparts without DNNs.

Grounded on the success of the second approach, we take the traditional DCT-based image coding method as the baseline, and we introduce a new module for sign compression, which is implemented and optimized as a DNN.

\begin{table}[!t]
\caption{Notations and definitions.}
\vspace{-5mm}
\label{tab:notation}
\begin{center}
\small
\begin{tabular}{c|l}
\bhline
\multicolumn{1}{c|}{Notation} & 
\multicolumn{1}{c}{Definition}\\\hline
$I, J$ & Width and height of image\\\hline
$[\,\cdot\,]_{i,j}$ 
& $(i,j)^\mathrm{th}$ value of image,\\
& $i, j$: horizontal and vertical indices\\\hline
$f_{i,j,\omega_i,\omega_j}$
& Unitary 2D Fourier basis function\\
& $\sim \exp{(-\sqrt{-1}\, 2\pi (i\omega_i /I + j\omega_j /J))}$,\\
& $\omega_i, \omega_j$: frequency indices along $i, j$ directions\\\hline
$U, V$ & Width and height of block\\\hline
$[\,\cdot\,]_{u,v}^{m,n}$
& $(u,v)^\mathrm{th}$ value of $(m,n)^\mathrm{th}$ block of image,\\
& $u, v$: indices in block, $m,n$: block indices\\\hline
$c_{u,v,\omega_u,\omega_v}$
& Orthonormal 2D cosine basis function \\
& $\sim \cos{\left((u+1/2)\omega_u\pi/U\right)}\cos{\left((v+1/2)\omega_v\pi/V\right)}$,\\
& $\omega_u, \omega_v$: frequency indices along $u, v$ directions\\\hline
$\alpha $
& Phase of complex variable \\\hline
$\phi_\theta$
& DNN parameterized by $\theta$ \\\hline
$\sum_{u, v} [\,\cdot\,]_{u, v}$
& Abridged notation of $\sum_{u=0}^{U-1}\sum_{v=0}^{V-1} [\,\cdot\,]_{u, v}$ \\\hline
$\sum_{m,n,u,v} [\,\cdot\,]_{u, v}^{m,n}$
& Sum of $\sum_{u,v} [\,\cdot\,]_{u, v}^{m,n}$ over all block $(m,n)$ \\\bhline
\end{tabular}
\end{center}
\end{table}

\section{Proposed method}
\label{s3}
In this section, we elaborate on how to compress the sign information of DCT coefficients. Our method is incorporated into classical still image coding. Specifically, we consider JPEG as an example, which consists only of basic components, i.e., DCT, quantization, and entropy coding. 

\subsection{Notations and definitions}
\label{s3ss1}
We introduce mathematical ingredients for clarity of explanation. Table~\ref{tab:notation} summarizes notations and definitions. Throughout this paper, we use Einstein notation~\cite{Einstein1916Die} to simply represent an image as well as its transformation in the mathematical form. In the context of Einstein notation, the summation over a specific index is simply represented with a single term, where the index appears twice.\vspace{2mm}

\subsubsection{Discrete Fourier transformation in Einstein notation}
\label{s3ss1sss1}
Let $x$ be a grayscale image of size $I \times J$. A single term $f_{i,j,\omega_i,\omega_j} x_{i,j}$ represents the 2D DFT operation
\begin{equation}
f_{i,j,\omega_i,\omega_j} x_{i,j}\sim\nonumber
\sum_{i,j} \exp\left(-\sqrt{-1} \, 2\pi \left(
\frac{i\omega_i}{I}+
\frac{j\omega_j}{J}\right)\right) 
x_{i,j},
\label{eq:fdft_full}
\end{equation}
where $i,\omega_i \in \{0, \ldots, I-1\}$, and $j, \omega_j \in \{0, \ldots, J-1\}$. Let $y_{\omega_i, \omega_j}$ be one of the DFT coefficients of $x$, such that 
\begin{equation}
y_{\omega_i, \omega_j} = f_{i,j,\omega_i,\omega_j} x_{i,j}.
\label{eq:fdft_ein}
\end{equation}
We simply write inverse DFT of $y$ as
\begin{equation}
x_{i,j} = \bar{f}_{i,j,\omega_i,\omega_j} y_{\omega_i,\omega_j},
\label{eq:idft_ein}
\end{equation}
where $\bar{f}$ is the complex conjugate of $f$.\vspace{2mm}
\subsubsection{Discrete cosine transformation in Einstein notation}
\label{s3ss1sss2}
Let $x^{m,n}$ be the $(m,n)^\mathrm{th}$ block of size $U \times V$ extracted from a grayscale image $x$. We denote 2D DCT in Einstein notation: 
\begin{equation}
c_{u,v,\omega_u,\omega_v} x_{u,v}^{m,n} \sim
\sum_{u,v} 
\cos\frac{(u+1/2)\omega_u\pi}{U}
\cos\frac{(v+1/2)\omega_v\pi}{V} 
x_{u,v}^{m,n},
\label{eq:fdct_full}
\end{equation}
where $u,\omega_{u} \in \{0, \cdots, U-1\}$, and $v, \omega_{v} \in \{0, \dots, V-1\}$. Let $y_{\omega_{u},\omega_{v}}^{m,n}$ be one of the DCT coefficients of $x^{m,n}$, such that 
\begin{equation}
y_{\omega_{u},\omega_{v}}^{m,n}=
c_{u,v,\omega_{u},\omega_{v}}^{} x_{u,v}^{m,n}.
\end{equation}
We also write inverse DCT as
\begin{equation}
x_{u,v}^{m,n} =
c_{u,v,\omega_{u},\omega_{v}} y_{\omega_{u},\omega_{v}}^{m,n}.
\end{equation}

\subsection{Encoder and decoder}
\label{s3ss2}
Our method is incorporated in a conventional DCT-based image coding method, i.e., JPEG. As shown in Fig.~\ref{fig:encdec}, we extract the signs of the quantized DCT coefficients and compress them through sign retrieval. We also modify the standard bitstream of JPEG to describe the magnitudes and the compressed signs separately. Our method consists of an encoder and a decoder as described below.\vspace{2mm}

\noindent
\textbf{Encoder}: We first separate quantized DCT coefficients $y_{\omega_{u},\omega_{v}}^{m,n}$ into the magnitudes $|y_{\omega_{u},\omega_{v}}^{m,n}|$ and signs $\mathrm{sign}(y_{\omega_{u},\omega_{v}}^{m,n})$. The magnitudes are compressed using suitable entropy coding, e.g., \cite{Huffman1952method,Golomb1966run,Rice1971adaptive,Martin1979range}. However, we compress the signs through sign retrieval, which will be detailed in Sections~\ref{s3ss3}, \ref{s3ss4}, and \ref{s3ss5}. To summarize, we restore the signs only from the magnitudes $|y_{\omega_{u},\omega_{v}}^{m,n}|$ without using the original signs. Finally, since the restored signs may be incorrect, we append the residual information
\begin{equation}
e_{\omega_{u},\omega_{v}}^{m,n}=
\mathrm{sign}{(y_{\omega_{u},\omega_{v}}^{m,n})} \oplus \mathrm{sign}{(c_{u,v,\omega_{u},\omega_{v}}\tilde{x}_{u,v}^{m,n})}
\label{eq:res}
\end{equation}
to the bitstream, where $\oplus$ is the XOR operator, and $\tilde{x}$ is an image restored by our sign retrieval algorithm. In this way, the signs are represented as the residual $e_{\omega_{u},\omega_{v}}^{m,n}$; all the signs can be recovered without loss from the residual. The performance of our method depends on the code length for the residual. In this paper, we compute the minimum code length from the entropy of the residual. In real applications, the actual code length will be determined by several factors, such as the method for entropy coding and the data unit size for the bitstream.\vspace{2mm}

\noindent
\textbf{Remarks on encoder}: If the signs are restored correctly to some extent, the residual has many zeros but few ones. In this case, the residual is compressible using entropy coding methods~\cite{Reininger1983distribution,Bellifemine1992statistical,Lam2000math}. Therefore, the accuracy of sign retrieval is essential for sign compression.\vspace{2mm}

\noindent
\textbf{Decoder}: The decoder first parses the bitstream to obtain the magnitudes $|y_{\omega_{u},\omega_{v}}^{m,n}|$ and the residual $e_{\omega_{u},\omega_{v}}^{m,n}$. The corresponding signs are then restored only from the magnitudes by our sign retrieval algorithm in the same manner as in the encoder. The restored signs including errors are corrected by $e_{\omega_{u},\omega_{v}}^{m,n}$. Merging the magnitudes with the correct signs, we reproduce the DCT coefficients $y_{\omega_{u},\omega_{v}}^{m,n}$. We finally obtain a decoded image by inverse DCT of the reproduced coefficients.\vspace{2mm}

\noindent
\textbf{Remarks on decoder}: Because the signs restored by the sign retrieval algorithm are finally corrected by the residual, the DCT coefficients $y_{\omega_{u},\omega_{v}}^{m,n}$ in our decoder are exactly the same as those in the baseline JPEG. Consequently, our decoded image is also the same as that obtained with JPEG; there are no differences between these images. We will thus evaluate our method in terms of the bit amount required for the sign residual, which is what we need to losslessly transmit the original sign information, i.e., the residual information $e_{\omega_{u},\omega_{v}}^{m,n}$ for all $m$, $n$, $\omega_u$, and $\omega_v$.

\begin{figure}[!t]
\centering
\begin{minipage}[!t]{0.49\linewidth}
\centering
\includegraphics[width=70mm]{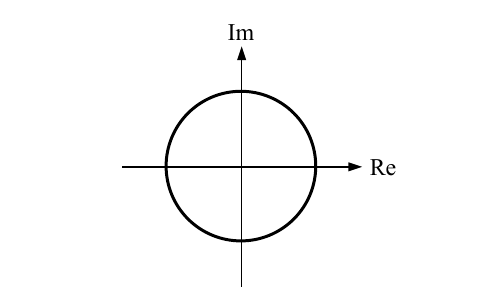}
\subcaption{Circle for \eqref{eq:pr_dft_0}}
\end{minipage}
\begin{minipage}[!t]{0.49\linewidth}
\centering
\includegraphics[width=70mm]{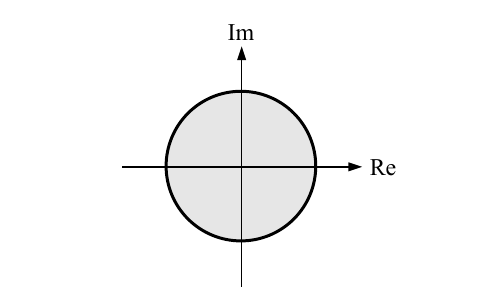}
\subcaption{Disk for \eqref{eq:pr_dft_1}}
\end{minipage}
\caption{Solution spaces in DFT domain.}
\label{fig:const_dft}
\end{figure}

\subsection{Formulation of sign retrieval}
\label{s3ss3}
The phase retrieval problem~\cite{Gerchberg1972practical,Fienup1982phase,Fickus14phase,Candes2015phaseA,Candes2015phaseB,Goldstein2018phasemax,Bahmani2017flexible,Salehi18learning} aims to solve a non-linear system of equations with respect to an image $x$,
\begin{equation}
|y_{\omega_i,\omega_j}| = |f_{i,j,\omega_i,\omega_j} x_{i,j}|, \quad \forall \omega_i, \omega_j,
\label{eq:pr_dft_0}
\end{equation}
where $f_{i,j,\omega_i,\omega_j}$ is the DFT basis. Symbol $y_{\omega_i,\omega_j}$ represents the DFT coefficient for each $(\omega_i,\omega_j)$, but only its magnitude is known in the phase retrieval problem. In other words, we have to find the true phase information of $y_{\omega_i,\omega_j}$ using only the known magnitude $|y_{\omega_i,\omega_j}|$. However, we cannot find a unique solution for this. The closed-form solution to \eqref{eq:pr_dft_0} is
\begin{equation}
\left\{
\begin{aligned}
x_{i,j}&=\bar{f}_{i,j,\omega_i,\omega_j}z_{\omega_i,\omega_j}\\
z_{\omega_i,\omega_j}&=
\exp(\sqrt{-1} \alpha_{\omega_i,\omega_j} ) |y_{\omega_i,\omega_j}|,\,\,
\alpha_{\omega_i,\omega_j} \in [0, 2\pi)
\end{aligned}
\right.,
\end{equation}
where $\alpha_{\omega_i,\omega_j}$ is an arbitrary phase and defined for each of the frequencies $(\omega_i,\omega_j)$. Therefore, \eqref{eq:pr_dft_0} has an infinite number of solutions.

To avoid this difficulty, the authors of \cite{Bahmani2017flexible,Goldstein2018phasemax,Salehi18learning} proposed a convex relaxation of \eqref{eq:pr_dft_0} with regularization, written as,
\begin{equation}
\min_{\tilde{x}}\quad
L(\tilde{x}) \quad \mathrm{s.t.}\quad
|y_{\omega_i,\omega_j}| \geq 
|f_{i,j,\omega_i,\omega_j} \tilde{x}_{i,j}|, \quad \forall \omega_i, \omega_j,
\label{eq:pr_dft_1}
\end{equation}
where $L(\tilde{x})$ is an objective function with respect to the restored image $\tilde{x}$. With a suitable definition of $L(\tilde{x})$, a unique solution can be found and is equivalent to the correct solution, i.e., $\tilde{x}=x$. For example, the authors of \cite{Bahmani2017flexible,Goldstein2018phasemax} defined $L(\tilde{x})$ as the inner product function between $\tilde{x}$ and the so-called anchor vector.

Figure~\ref{fig:const_dft} shows a geometrical interpretation for the convex relaxation in the DFT domain; the solution spaces for \eqref{eq:pr_dft_0} and \eqref{eq:pr_dft_1} are illustrated in (a) and (b), respectively. The solution is limited to the circle in \eqref{eq:pr_dft_0}, while the solution space is extended to the disk (the area inside the circle) in \eqref{eq:pr_dft_1}. Note that the solution space is non-convex for the circle, but it is convex for the disk. Therefore, the solution to \eqref{eq:pr_dft_1} can be efficiently computed by a convex optimization algorithm.

In our sign retrieval, we regard the sign as a special case of the phase. Because DCT coefficients are always real values without imaginary parts, their phases are limited to $0$ or $\pi$. In this case, the function $\exp(\sqrt{-1} \alpha_{\omega_u,\omega_v}^{m,n})$ takes only $1$ or $-1$, which is equivalent to the signs of real values:
\begin{equation}
\mathrm{sign}(y_{\omega_u,\omega_v}^{m,n})=
\exp(\sqrt{-1} \alpha_{\omega_u,\omega_v}^{m,n}), \quad \alpha_{\omega_u,\omega_v}^{m,n} \in \{0, \pi\}.
\label{eq:sign_phase}
\end{equation}
Accordingly, the phase restoration in the complex domain reduces to the sign restoration in the real domain.

\begin{figure}[!t]
\centering
\begin{minipage}[!t]{0.49\linewidth}
\centering
\includegraphics[width=70mm]{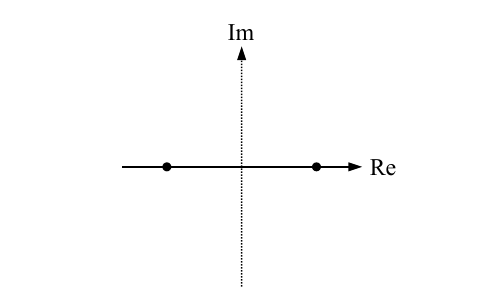}
\subcaption{Two points for \eqref{eq:pr_dct_0}}
\end{minipage}
\begin{minipage}[!t]{0.49\linewidth}
\centering
\includegraphics[width=70mm]{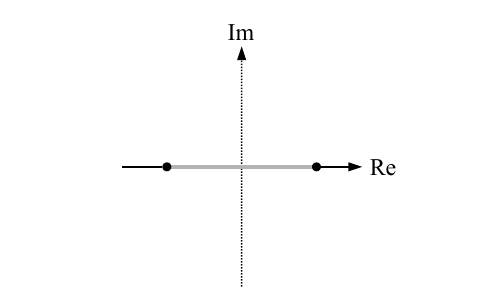}
\subcaption{Line for \eqref{eq:pr_dct_1}}
\end{minipage}
\caption{Solution spaces in DCT domain.}
\label{fig:const_dct}
\end{figure}

The sign restoration in the DCT domain is formulated as follows. Replacing $f_{i,j,\omega_i,\omega_j}$ and $|y_{\omega_i,\omega_j}|$ in \eqref{eq:pr_dft_0} with $c_{u,v,\omega_u,\omega_v}$ and $|y_{\omega_u,\omega_v}^{m,n}|$, respectively, we obtain 
\begin{equation}
|y_{\omega_{u},\omega_{v}}^{m,n}| =
|c_{u,v,\omega_{u},\omega_{v}} x_{u,v}^{m,n}|, \quad \forall m,n,\omega_u, \omega_v.
\label{eq:pr_dct_0}
\end{equation}
Following \cite{Bahmani2017flexible,Goldstein2018phasemax,Salehi18learning}, we then introduce a convex relaxation for \eqref{eq:pr_dct_0} as
\begin{equation}
\min_{\tilde{x}}\quad
L(\tilde{x})
\quad \mathrm{s.t.}\quad
|y_{\omega_{u},\omega_{v}}^{m,n}| \geq 
|c_{u,v,\omega_{u},\omega_{v}} \tilde{x}_{u,v}^{m,n}|,
\quad \forall  m,n, \omega_u, \omega_v.
\label{eq:pr_dct_1}
\end{equation}
Figure~\ref{fig:const_dct} shows the solution spaces for \eqref{eq:pr_dct_0} and \eqref{eq:pr_dct_1}. The solution space is limited to the two isolated points for \eqref{eq:pr_dct_0}, while it is extended to the line connecting the two points for \eqref{eq:pr_dct_1}. 

We implicitly define the objective function for \eqref{eq:pr_dct_1} by using a DNN. Let $\phi_\theta$ be a DNN parameterized by $\theta$, and let $x_0$ be an initial image given as the input to the DNN. The loss function with the pre-trained DNN is described as 
\begin{equation}
L(\tilde{x};\theta) = 
\sum_{m,n,u,v}\,\,
(\tilde{x}_{u,v}^{m,n} - \phi_{\theta}(x_0)_{u,v}^{m,n})^2,
\label{eq:pr_dct_1.5}
\end{equation}
where $\phi_{\theta}(x_0)_{u,v}^{m,n}$ represents the $(u,v)^\mathrm{th}$ pixel value in the $(m,n)^\mathrm{th}$ block of the image $\phi(x_0)$. The initial image $x_0$ is computed by inverse DCT of the DC components, which are already encoded and decoded before sign retrieval. Because $x_0$ is a band-limited version of $x$, a super resolution network would be suitable for the DNN $\phi_\theta$. The details of our network architecture will be given in Section~\ref{s3ss5}.

The objective function in \eqref{eq:pr_dct_1.5} is conditioned on a set of images used for training the DNN. To train the DNN, we optimize the parameter $\theta$ over the training images $X$ as
\begin{equation}
\min_{\theta} \,\,\,
\sum_{x \in X} L(x;\theta)
\quad \mathrm{s.t.}\quad
|y_{\omega_{u},\omega_{v}}^{m,n}| \geq 
|c_{u,v,\omega_{u},\omega_{v}} \phi_{\theta}(x_0)_{u,v}^{m,n}|,
\quad \forall m,n,\omega_u, \omega_v,
\label{eq:pr_dct_2}
\end{equation}
where $ |y_{\omega_{u},\omega_{v}}^{m,n}|$ and $x_0$ are computed from the corresponding image $x$ in the training dataset $X$.

\subsection{Projection onto convex sets based solution}
\label{s3ss4}
We derive our solution method for \eqref{eq:pr_dct_1} and \eqref{eq:pr_dct_2} from the framework of POCS~\cite{Bauschke1996projection}. Note that \eqref{eq:pr_dct_2} describes the training process of the DNN over the training images $X$, while \eqref{eq:pr_dct_1} describes the inference process for a target image $\tilde{x}$ using the pre-trained DNN $\phi_\theta$.

\begin{figure}[!t]
\centering
\includegraphics[width=140mm]{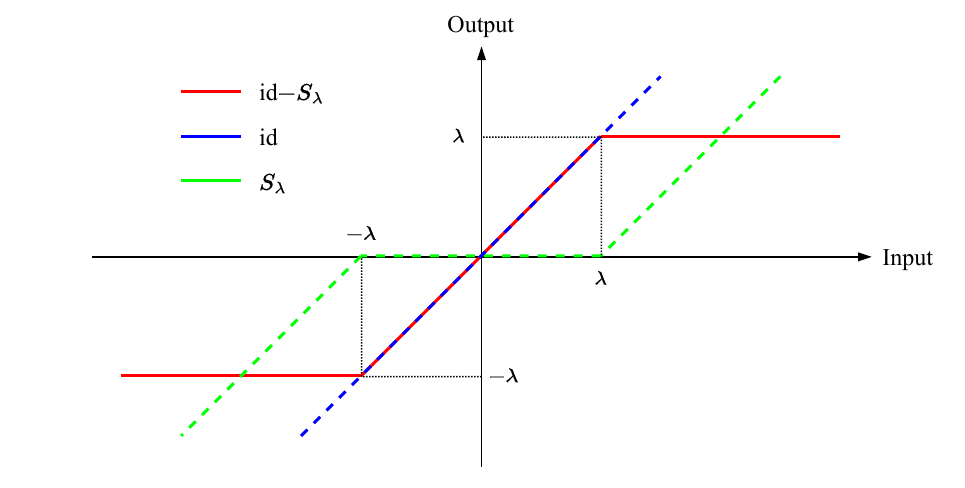}
\caption{$\mathrm{id}-\mathcal{S}_\lambda$, $\mathrm{id}$, and $\mathcal{S}_\lambda$.}
\label{fig:proj}
\end{figure}

We first consider the constraint in \eqref{eq:pr_dct_1} and \eqref{eq:pr_dct_2}. Let $\tilde{X}$ be the solution space satisfying the constraint, given by
\begin{equation}
\tilde{X} = \{\tilde{x} : |y_{\omega_{u},\omega_{v}}^{m,n}| \geq 
|c_{u,v,\omega_{u},\omega_{v}} \tilde{x}_{u,v}^{m,n}|, \forall m,n,\omega_u, \omega_v\}.
\label{eq:set}
\end{equation}
We then define the following projection operator:
\begin{equation}
\mathrm{proj}_{\tilde{X}}(z) = 
\mathop{\mathrm{argmin}}_{\tilde{x} \in \tilde{X}}\,\,
\sum_{m,n,u,v}\,\,(z_{u,v}^{m,n}-\tilde{x}_{u,v}^{m,n})^2
\label{eq:pocs}
\end{equation}
whose solution satisfies the constraint $\tilde{x} \in\tilde{X}$ while minimizing the squared distance between the input $z$ and the solution. Thanks to Parseval's theorem~\cite{Hardy1931note}, i.e., the orthonomality of DCT, the objective function of \eqref{eq:pocs} can be minimized in the DCT domain because the squared distance in the DCT domain equals that in the image domain. The closed-form solution to \eqref{eq:pocs} is given as
\begin{equation}
\mathrm{proj}_{\tilde{X}}(z)_{u,v}^{m,n}=
c_{u,v,\omega_u,\omega_v}
(\mathrm{id}-\mathcal{S}_{\lambda})(c_{u,v,\omega_u,\omega_v}z_{u,v}^{m,n}),
\label{eq:solution}
\end{equation}
where $\mathrm{id}$ is an identical function, and $\mathcal{S}_\lambda$ is a soft thresholding function~\cite{Donoho1995denoising} with the threshold value $\lambda=|y_{\omega_{u},\omega_{v}}^{m,n}|$. The function $\mathrm{id}-\mathcal{S}_\lambda$ applies POCS in the DCT domain; if an input DCT coefficient is outside the range $[-\lambda,\lambda]$, POCS forces the output to be $\pm \lambda$, as shown in Fig.~\ref{fig:proj}.

Given the closed-form solution in \eqref{eq:solution}, we can simplify \eqref{eq:pr_dct_1} and \eqref{eq:pr_dct_2}. We denote the composite mapping of $\mathrm{proj}_{\tilde{X}}$ and $\phi_\theta$ as $\psi_\theta$;
\begin{equation}
\psi_\theta(x_0) = \mathrm{proj}_{\tilde{X}} \Big(\phi_\theta(x_0)\Big) =
\mathrm{proj}_{\tilde{X}} \circ \phi_\theta(x_0).
\label{eq:comp}
\end{equation}
The operator $\circ$ means composition of two functions; $\mathrm{proj}_{\tilde{X}} \circ \phi_\theta(x_0)$ is equivalent to $ \mathrm{proj}_{\tilde{X}} (\phi_\theta(x_0))$.
%
We can remove the constraints from \eqref{eq:pr_dct_1} and \eqref{eq:pr_dct_2} and rewrite them respectively as 
\begin{equation}
\min_{\tilde{x}}
\sum_{m,n,u,v}\,\,
(\tilde{x}_{u,v}^{m,n} - \psi_\theta(x_0)_{u,v}^{m,n})^2
\label{eq:pr_dct_3}
\end{equation}
\begin{equation}
\min_{\theta} \quad 
\sum_{x \in X}\sum_{m,n,u,v}\,\,
(x_{u,v}^{m,n} - \psi_\theta(x_0)_{u,v}^{m,n})^2.
\label{eq:pr_dct_4}
\end{equation}
The solution for \eqref{eq:pr_dct_3} is obviously $\tilde{x}=\psi_\theta(x_0)$, which can be computed by forward inference on the pre-trained DNN and POCS. Meanwhile, minimization of \eqref{eq:pr_dct_4} is not that straightforward. Fortunately, \eqref{eq:solution} is sub-differentiable~\cite{Boyd2004convex} as shown in Fig.~\ref{fig:proj}, and the entire objective function of \eqref{eq:pr_dct_4} is also sub-differentiable. We can thus solve \eqref{eq:pr_dct_4} using optimizers such as sub-gradient descent~\cite{Duchi2011adaptive} or gradient descent~\cite{Kingma2015adam}. We use automatic differentiation methods~\cite{Baydin2017automatic} implemented in machine learning frameworks~\cite{Pedregosa2011scikit,Abadi2016tensorflow,Paszke2019pythorch}.\vspace{2mm}

\begin{table}[!t]
\caption{Network architecture of $\phi_\theta$.}
\vspace{-5mm}
\label{tab:network}
\begin{center}
\small
\renewcommand{\arraystretch}{1.2}
\begin{tabular}{wc{12mm}ccccc}
\bhline
\multicolumn{1}{c}{Layer} & 
\multicolumn{1}{c}{Ker. size} & 
\multicolumn{1}{c}{In ch.} & 
\multicolumn{1}{c}{Out ch.} &
\multicolumn{1}{c}{Act.} &
\multicolumn{1}{c}{Input}\\\hline
Conv-1  & $5 \times 5$ &   $1$ & $64$ & ReLU & Image  \\
Conv-2  & $1 \times 1$ &  $64$ & $32$ & ReLU & Conv-1 \\
Conv-3  & $3 \times 3$ &  $32$ &  $1$ &   -- & Conv-2 \\\bhline
\end{tabular}
\end{center}
\end{table}

\begin{table}[!t]
\caption{Computing environment.}
\vspace{-5mm}
\label{tab:comp}
\begin{center}
\small
\renewcommand{\arraystretch}{1.2}
\begin{tabular}{wc{33mm}|c}  \bhline
CPU & Intel Core i9-10900K\\ 
Main memory & 64~GB \\ 
GPU & NVIDIA GeForce RTX3090 \\ 
GPU memory  & 24~GB \\ 
OS  & Ubuntu 20.04 LTS \\ 
Language \& framework & Python 3.8.10 \& PyTorch 1.11.0\\\bhline
\end{tabular}
\end{center}
\end{table}

\noindent
\textbf{Remarks on POCS}: The derivation of \eqref{eq:solution} is based on the orthonormality of DCT. However, other transformations, such as the integer DCTs~\cite{Chan1991order,Chen2000video,Malvar2003low,Meher2014efficient,Zhao2021transform} adopted in modern video coding standards~\cite{Sullivan2012overview,Bross2021overview}, are not orthonormal. Since we do not have closed-form solutions for these transformations, we might resort to other convex optimization techniques~\cite{Adler2018learned} to solve \eqref{eq:pr_dct_4}, which we leave as future work.

\subsection{Implementation of deep neural network}
\label{s3ss5}
We use a trainable DNN $\phi_\theta$ as a component of the objective function as shown in \eqref{eq:pr_dct_1.5}. We describe the design choice for $\phi_\theta$, which is interpreted as a super-resolution network.

In \cite{Dong2016image,Shi2016real,Kim2016accurate,Ledig2017photo,Tai2017image,Yang2020learning,Anwar2020densely,Wang2021unsupervised}, various super resolution methods were proposed, most of which aimed to improve the quality of super-resolved images using a large number of network parameters. However, a lower number of network parameters is preferable to adapt to various image coding scenarios, such as those involving on memory-limited devices like GPUs. We thus use the simplest super resolution network proposed by Dong et al.~\cite{Dong2016image}, whose architecture is summarized in Table~\ref{tab:network}. Sign retrieval using this network is referred to as ``prototype deep sign retrieval (PDSR).''

\begin{figure}[!t]
\centering
\begin{minipage}[!t]{0.99\linewidth}
\centering
\hspace{-13mm}
\includegraphics[width=85mm]{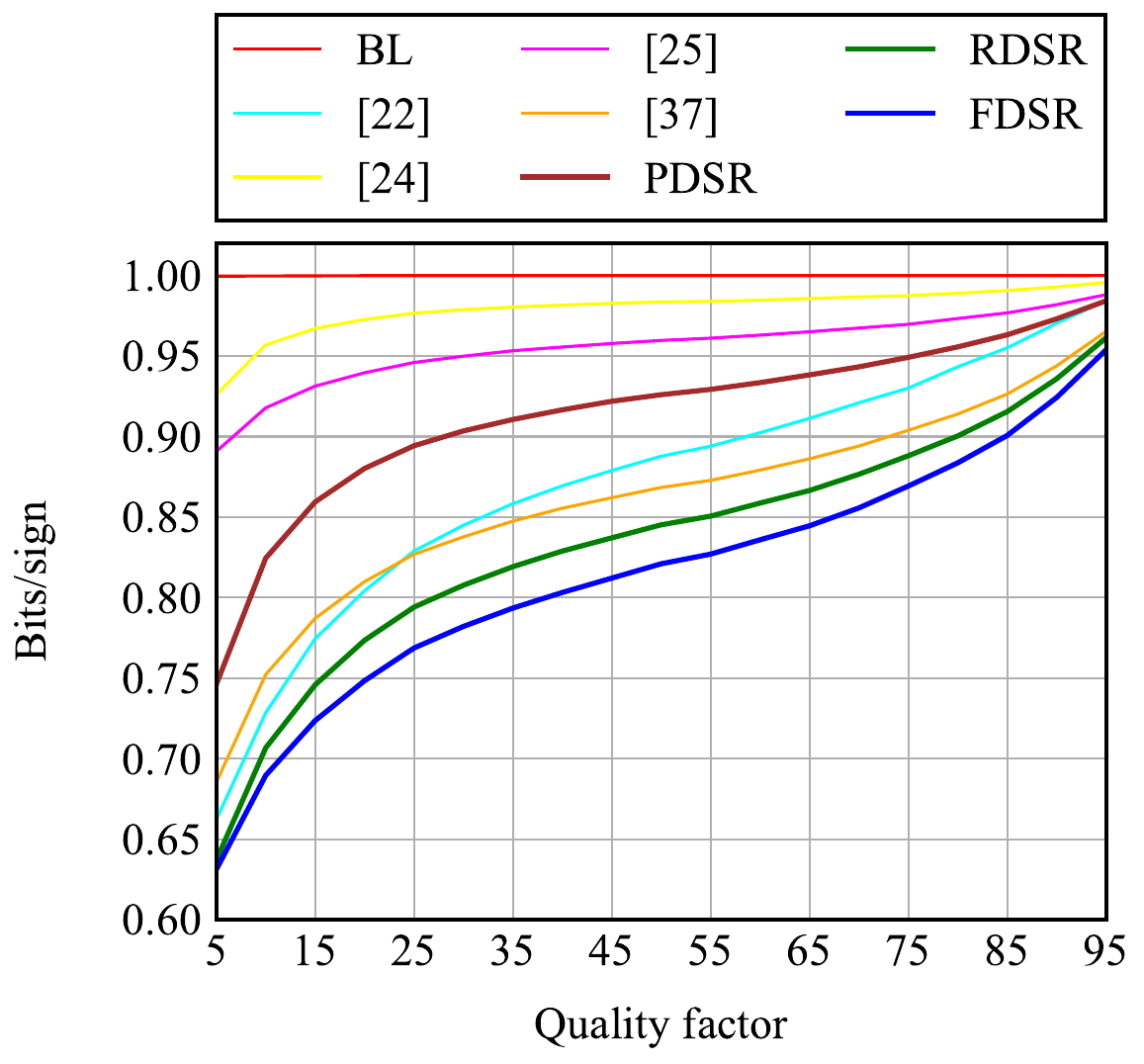}
\subcaption{BPS}
\end{minipage}\vspace{2mm}
\begin{minipage}[!t]{0.99\linewidth}
\centering
\hspace{-13mm}
\includegraphics[width=85mm]{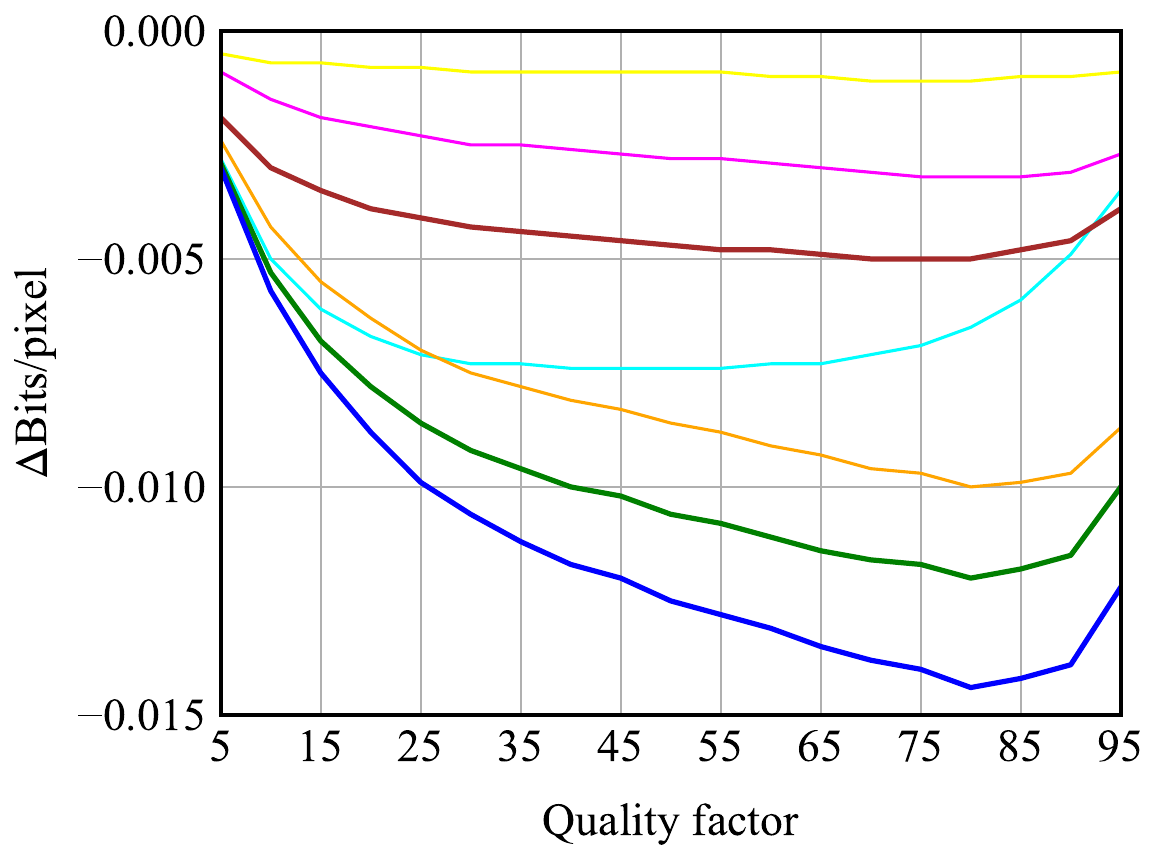}
\subcaption{$\Delta$BPP}
\end{minipage}
\caption{BPS and BPP of previous and our methods (PDSR, RDSR, and FDSR).}
\label{fig:bpsbpp}
\end{figure}

We then tailor another network with the same number of parameters as PDSR. Our design here is motivated by the recursive architecture~\cite{Tai2017image}, where a basic network with shared parameters is used recursively to gradually improve the quality of super-resolved images. Sign retrieval using this recursive architecture is called ``recursive deep sign retrieval (RDSR),'' which is described as the $K$-times repetition of PDSR:
\begin{equation}
\psi_\theta(x_0)=
\underbrace{\mathrm{proj}_{\tilde{X}} \circ \phi_\theta \circ \cdots \circ \mathrm{proj}_{\tilde{X}} \circ \phi_\theta}_{K \,\text{times}}(x_0).
\end{equation}
Note that the parameter $\theta$ is shared among all the elemental networks $\phi_\theta$. RDSR is the complete implementation of our method for sign retrieval, which is illustrated in Fig.~\ref{fig:sr}. 

To investigate the upper-bound performance of RDSR, we also develop ``full deep sign retrieval (FDSR),'' where the parameter sharing in RDSR is nullified:
\begin{equation}
\psi_\theta(x_0)=
\mathrm{proj}_{\tilde{X}} \circ 
\phi_{\theta_{K-1}} \circ \cdots \circ \mathrm{proj}_{\tilde{X}} \circ \phi_{\theta_0}(x_0).
\end{equation}
Here, $\theta_{k}$ is a distinct set of parameters for the $k$-th network. Using $K$-times more parameters than RDSR, FDSR is expected to achieve better quality for sign retrieval. However, as will be demonstrated in the next section, the gain of FDSR over RDSR is not significant. We argue that RDSR is more preferable than FDSR due to the good balance between the performance and the number of network parameters.

We finally remark in this section that all the methods mentioned here (PDSR, RDSR, and FDSR) restore the images rather than the signs themselves. The restored images are then used to extract the restored signs via forward DCT as illustrated in Fig.~\ref{fig:sr}.

\section{Experiment}
\label{s4}
\subsection{Configuration}
\label{s4ss1}
We conducted our experiments on the computing environment in Table~\ref{tab:comp}. We used grayscale images in all of the experiments. We used a block size of $U \times V = 8 \times 8$ in the training and test phases. PDSR, RDSR, and FDSR were trained on $50,000$ image patches with a size of $I \times J = 256 \times 256$, and they were randomly cropped from $585$ images included in CLIC2021\footnote{\url{http://clic.compression.cc/2021/}}. These images were written as $X$ in \eqref{eq:pr_dct_4}; these uncompressed images were the ground truth for the loss function. We applied block-wise DCT to all the original image patches, and we quantized the DCT coefficients with the quality factor $\text{QF}=50$. We used the following quantization table.
\[
\left[
\begin{array}{rrrrrrrr}
\small
16 & 11 & 10 & 16 & 24  & 40  & 51  & 61  \\
12 & 12 & 14 & 19 & 26  & 58  & 60  & 55  \\
14 & 13 & 16 & 24 & 40  & 57  & 69  & 56  \\
14 & 17 & 22 & 29 & 51  & 87  & 80  & 62  \\
18 & 22 & 37 & 56 & 68  & 109 & 103 & 77  \\
24 & 35 & 55 & 64 & 81  & 104 & 113 & 92  \\
49 & 64 & 78 & 87 & 103 & 121 & 120 & 101 \\
72 & 92 & 95 & 98 & 112 & 100 & 103 & 99
\end{array}
\right]
\nonumber
\]
The quantization process was the same as that in JPEG~\cite{Wallace1991jpeg}. We used TorchJPEG~\cite{Ehrlich2020quantization} to implement DCT and quantization. The quantized DCT coefficients for each of the patches were used to determine the thresholds $\lambda$ in \eqref{eq:solution}. We set the number of networks $K$ to $20$ for RDSR and FDSR. The optimizer was Adam~\cite{Kingma2015adam} with the learning rate $2\mathrm{E}-4$. The batch size and the number of epochs were $10$ and $50$, respectively. 

\begin{table}[!t]
\caption{Reduction rate of BPS against that of baseline.}
\vspace{-5mm}
\label{tab:bps}
\begin{center}
\small
\renewcommand{\arraystretch}{1.2}
\begin{tabular}{c|c|c|c|c|c|c|c} \bhline
&  \cite{Nakagawa2017sign} & \cite{Tsutake21efficient} & \cite{Tu2002context} &  \cite{Miroshnichenko18compression}  & PDSR & RDSR & FDSR\\\hline
Lowest  & 0.02 & 0.03 & 0.00 & 0.01 & 0.02 & 0.04 & 0.05 \\ 
Highest & 0.34 & 0.31 & 0.07 & 0.11 & 0.25 & 0.36 & 0.37 \\ 
Mean& 0.13 & 0.14 & 0.02 & 0.04 & 0.09 & 0.17 & 0.19 \\ \bhline
\end{tabular}
\end{center}
\end{table}

\begin{table}[!t]
\caption{Reduction rate of BPP against that of baseline.}
\vspace{-5mm}
\label{tab:bpp}
\begin{center}
\small
\renewcommand{\arraystretch}{1.2}
\begin{tabular}{c|c|c|c|c|c|c|c} \bhline
&  \cite{Nakagawa2017sign} & \cite{Tsutake21efficient} & \cite{Tu2002context} &  \cite{Miroshnichenko18compression}  & PDSR & RDSR & FDSR\\\hline
Lowest  & 0.01 & 0.03 & 0.00 & 0.01 & 0.01 & 0.03 & 0.04\\ 
Highest & 0.30 & 0.26 & 0.05 & 0.10 & 0.20 & 0.31 & 0.37\\ 
Mean& 0.10 & 0.11 & 0.01 & 0.04 & 0.06 & 0.13 & 0.15\\\bhline
\end{tabular}
\end{center}
\end{table}

For evaluation, we sampled $60$ images not included in the training dataset from CLIC2021. These test images were quantized with $\text{QF}=5, 10, \cdots, 95$ in the DCT domain. The signs of these DCT coefficients were compressed by using our method (PDSR, RDSR, and FDSR), a prediction-based method~\cite{Nakagawa2017sign}, and context-based ones~\cite{Tu2002context,Miroshnichenko18compression}. We also evaluated our previous method (sign retrieval without a DNN)~\cite{Tsutake21efficient}\footnote{The main difference between \cite{Tsutake21efficient} and this study is the definition of the loss function $L(\tilde{x})$ and a solution method for \eqref{eq:pr_dct_1}. In \cite{Tsutake21efficient}, we first defined the loss function as the $\ell_1$-norm of the wavelet transformation of $\tilde{x}$. The optimization problem, i.e., \eqref{eq:pr_dct_1}, was then solved by a phase-retrieval-based iterative method, referred to as the Fienup method~\cite{Fienup1982phase} (also known as the alternating minimization method~\cite{Netrapalli2013pahse}). The computation cost of the Fienup method in \cite{Tsutake21efficient} was very large because a large number of iterations was required to restore $\tilde{x}$.} to investigate how much the performance can be improved by using a DNN. All the methods were embedded into the standard JPEG pipeline, where only the sign compression part is different method to method.

\subsection{Compression efficiency}
\label{s4ss2}
All the methods compared in this paper comply with our coding scenario; the original signs should be reconstructed in the decoders without any loss, resulting in exactly the same decoded image as the one obtained with the standard JPEG. Therefore, only the rate performance can be compared among them. We here use two metrics for this: bits per sign~(BPS) and bits per pixel~(BPP), where the number of bits required for the sign information is divided by the number of signs for BPS and the number of pixels for BPP. Since some of the methods (ours and \cite{Nakagawa2017sign,Tsutake21efficient}) transmit the prediction residual instead of the signs themselves, the BPS and BPP values were computed from the entropy of the residual. On the other hand, we derived the BPS and BPP for \cite{Tu2002context,Miroshnichenko18compression} from the conditional probability models of the original sign information. As the baseline, we also evaluated the case without sign compression, where the entropy of the original sign information was obtained without any conditional probability models.

Figure~\ref{fig:bpsbpp} illustrates BPS and BPP values (averaged over $60$ images) for the previous and our methods against various QFs. In (b), we show the differences of BPP ($\Delta$BPP) from the baseline to enhance the visibility; a lower $\Delta$BPP value means better compression performance. Tables~\ref{tab:bps} and \ref{tab:bpp} show relative reduction rates: how much the BPS and BPP values can be reduced from the baseline by the sign compression methods. PDSR significantly outperformed the context-based methods~\cite{Tu2002context,Miroshnichenko18compression}, but it is comparable to or less efficient than \cite{Nakagawa2017sign,Tsutake21efficient}. In comparison, FDSR and RDSR yielded the first and second best scores, thanks to the deeper network structure than that of PDSR. More specifically, FDSR and  RDSR reduced the BPS~(BPP) by $19$~($15$) and $17$~($13$) percent from the baseline. 

We mention the trade-off between the number of network parameters and the compression efficiency. RDSR has the same number of parameters ($4,033$) as PDSR, but RDSR drastically outperformed PDSR. Meanwhile, FDSR uses $20$ times the number of parameters ($80,660$) as RDSR, but the performance gain of FDSR over RDSR was marginal. We conclude from these results that the recursive architecture adopted in RDSR is effective for sign retrieval, while keeping the number of the parameters small.

\subsection{Computation cost}
\label{s4ss3}
For fair comparisons, we performed all the methods on the CPU in Table~\ref{tab:comp}.We measured the execution time for sign compression (including the computation of a BPS value) for a single image at a target QF; the execution time does not include the time for other operations in the standard JPEG pipeline, e.g., quantization. Table~\ref{table:time} shows the time averaged over all the QFs.

\begin{table}[!h]
\caption{Execution times~[s].}
\vspace{-5mm}
\label{table:time}
\begin{center}
\small
\renewcommand{\arraystretch}{1.2}
\begin{tabular}{c|c|c|c|c|c|c} \bhline
\cite{Nakagawa2017sign} & \cite{Tsutake21efficient} & \cite{Tu2002context} &  \cite{Miroshnichenko18compression} &  PDSR & RDSR & FDSR\\\hline
88.18 & 85.69 & $\,0.91\,$ & $\,6.33\,$ & 0.54  & 10.39 & 10.30 \\ \bhline
\end{tabular}
\end{center}
\vspace{-2mm}
\end{table}

\noindent
The context-based methods~\cite{Tu2002context,Miroshnichenko18compression} were fast because no prediction or restoration of the signs are required; only the conditional probability was computed. However, these methods were poor in terms of compression efficiency as mentioned earlier. The prediction/restoration-based methods without DNNs~\cite{Nakagawa2017sign,Tsutake21efficient} had prohibitively long computation times due to the iterative computation for minimizing the cost functions. In comparison, our methods (PDSR, RDSR, and FDSR) were significantly faster than \cite{Nakagawa2017sign,Tsutake21efficient}. This is due to the efficiency of the forward inference process on a pre-trained network. These results demonstrate that our methods are also advantageous in terms of computation cost.

\begin{figure}[!t]
\centering
\begin{minipage}[!t]{0.32\linewidth}
\centering
\includegraphics[width=35mm]{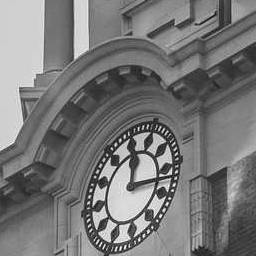}
\subcaption{Decoded}
\end{minipage}
\begin{minipage}[!t]{0.32\linewidth}
\centering
\includegraphics[width=35mm]{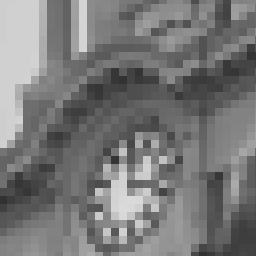}
\subcaption{Init, 19.78~dB}
\end{minipage}
\begin{minipage}[!t]{0.32\linewidth}
\centering
\includegraphics[width=35mm]{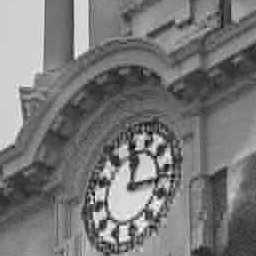}
\subcaption{\cite{Nakagawa2017sign}, 22.87~dB}
\end{minipage}\\\vspace{4mm}
\begin{minipage}[!t]{0.32\linewidth}
\centering
\includegraphics[width=35mm]{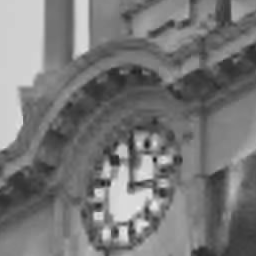}
\subcaption{\cite{Tsutake21efficient}, 21.58~dB}
\end{minipage}
\begin{minipage}[!t]{0.32\linewidth}
\centering
\includegraphics[width=35mm]{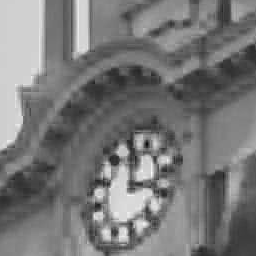}
\subcaption{PDSR, 21.29~dB}
\end{minipage}
\begin{minipage}[!t]{0.32\linewidth}
\centering
\includegraphics[width=35mm]{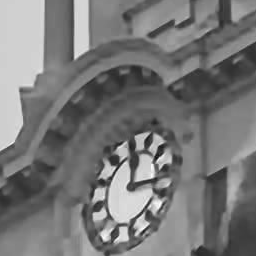}
\subcaption{RDSR, 23.21~dB}
\end{minipage}\\\vspace{4mm}
\begin{minipage}[!t]{0.32\linewidth}
\centering
\includegraphics[width=35mm]{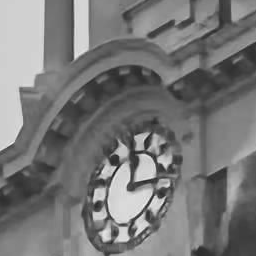}
\subcaption{FDSR, 23.28~dB}
\end{minipage}
\begin{minipage}[!t]{0.32\linewidth}
\centering
\includegraphics[width=35mm]{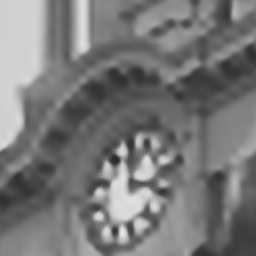}
\subcaption{RDSR w/o POCS, 20.79~dB}
\end{minipage}
\begin{minipage}[!t]{0.32\linewidth}
\centering
\includegraphics[width=35mm]{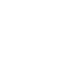}
\end{minipage}
\caption{Restored images.}
\label{fig:restored}
\end{figure}

\begin{figure}[!t]
\centering
\hspace{-13mm}
\includegraphics[width=85mm]{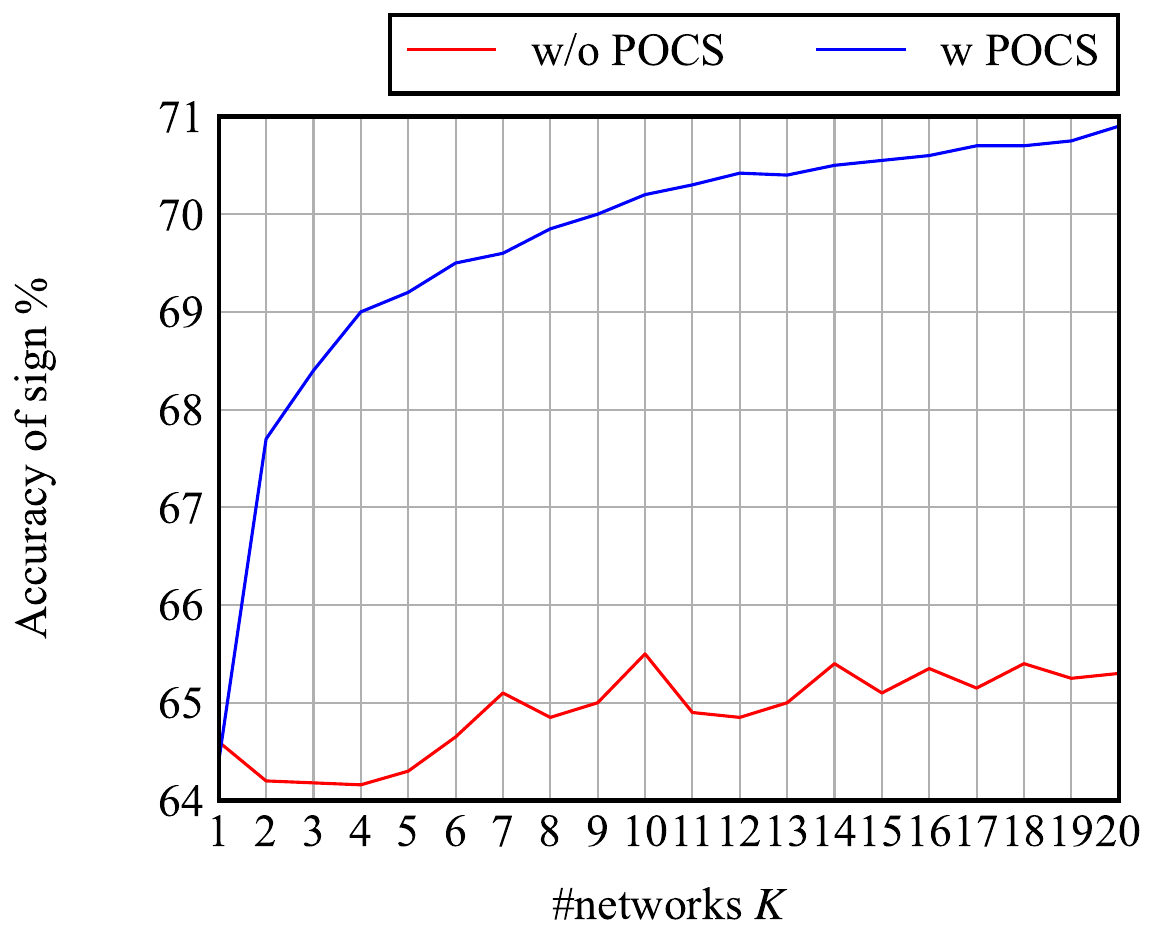}
\caption{Accuracy of sign against $K$.}
\label{fig:acc_k}
\end{figure}

\begin{figure}[!t]
\centering
\hspace{-13mm}
\includegraphics[width=85mm]{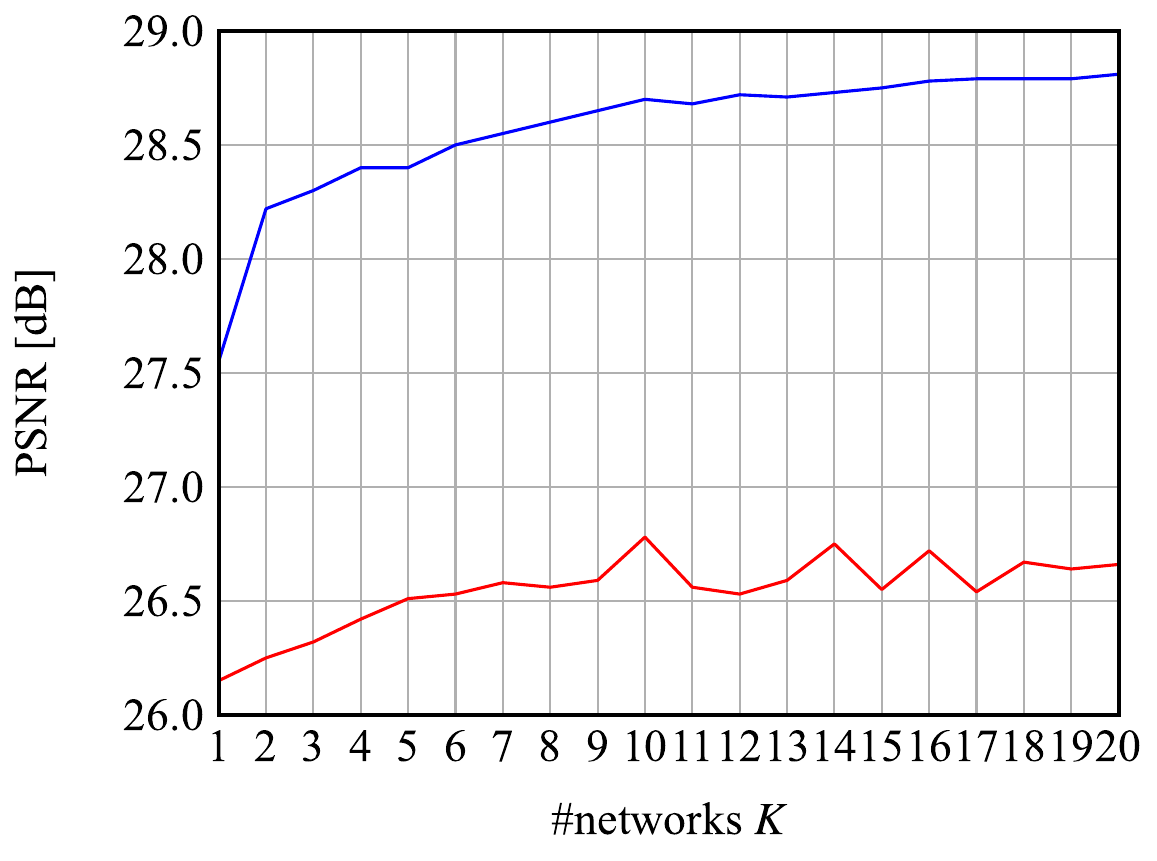}
\caption{PSNR values against $K$.}
\label{fig:psnr_k}
\end{figure}

\subsection{Restoration quality}
\label{s4ss4}
We hereafter investigate the reason behind the compression efficiency of RDSR and FDSR. To this end, we evaluated the quality of restored images before correcting the errors using the residual because better restoration quality leads to more efficient compression of the residual. Figure~\ref{fig:restored} shows (a) an image decoded by the standard JPEG, (b) the initial image (reconstructed from the DC components), and (c)--(g) the images restored by \cite{Nakagawa2017sign,Tsutake21efficient}, PDSR, RDSR, and FDSR, followed by (h) one without POCS (mentioned later). Note that these restored images are finally error-corrected using the residual to yield the same image as the decoded image (a). The peak signal-to-noise ratio~(PSNR) values reported in Fig.~\ref{fig:restored} were evaluated between the restored images and the decoded image.

We can observe from Fig.~\ref{fig:restored} that the previous methods~\cite{Nakagawa2017sign,Tsutake21efficient} and PDSR restored blurry textures and discontinuous edges. In comparison, RDSR restored sharp edges with little visible artifacts. It is worth noting that even though both PDSR and RDSR have the same number of parameters, i.e., $4,033$, the latter produced a significantly better image. Moreover, the difference between RDSR and FDSR is insignificant in terms of visual quality as well as PSNR value, indicating the effectiveness of the recursive architecture of RDSR.

\subsection{Role of projection onto convex sets}
\label{s4ss5}
We evaluate the role of POCS, one of the key elements in our method. We implemented a variant of RDSR, in which the POCS operator was replaced with the identical mapping; no operation was performed after each inference on a DNN $\phi_\theta$, which reduced RDSR to a naive super-resolution network. With various values for $K$, we trained RDSRs with and without POCS in the same configuration as that in Section~\ref{s4ss1}. As an evaluation metric, we defined the ``accuracy of sign (AoS),'' the accuracy of the restored signs with respect to the original signs, which equals the probability of zeros in the residual.

Figure~\ref{fig:acc_k} shows AoS curves against $K$, each of which is the average over the $60$ test images with $\text{QF}=50$. Figure~\ref{fig:psnr_k} shows average PSNR values against the ones decoded by the standard JPEG. We confirmed that without POCS, both the AoS and PSNR values did not increase with the increase of $K$. Meanwhile, with POCS, these values significantly increased with the increase of $K$. Figure~\ref{fig:restored}(h) shows an image restored without POCS at $K=20$; the difference between images (f) and (h) can solely be attributed to POCS. We conclude from these results that POCS played an important role in improving the restoration accuracy, resulting in better compression efficiency of the sign information in Section~\ref{s4ss2}.

\subsection{Effect of quantization factor}
\label{s4ss6}
Sign compression is performed after the quantization of the DCT coefficients. One may think that our method should be trained with various QFs. However, good generalization performance over a wide variety of QF values can be achieved with the training conditions described in Section~\ref{s4ss1}; we used an uncompressed image (without quantization) for the loss function and DCT coefficients quantized with $\text{QF}=50$ for POCS. To further investigate this issue, we performed several experiments with RDSR as follows.

We changed the training conditions with respect to the loss function; we quantized the training images with $\text{QF}=50$ in the DCT domain, and we used them as $X$ in the loss function of \eqref{eq:pr_dct_4}, while the QF for POCS was fixed to $50$. The average AoS value was $71$\% for the test time with $\text{QF}=50$. With the original training conditions, it was also $71$\%. Therefore, we conclude that quantization does not have a significant impact on the loss function. 

We also changed the training conditions with respect to the QFs for POCS. Specifically, we trained three networks with $\text{QF}=25$, $50$, and $75$ and tested them with $\text{QF}=25$, $50$, and $75$. The average AoS values depended only on the test QF; with $\text{QF}=25$, $50$, and $75$, the three networks resulted in the same average AoS values, $74$, $71$, and $68$ percents, respectively, regardless of the QFs at the training phases. We conclude that a single QF for POCS is sufficient for training our method.

\section{Conclusion}
\label{s5}
We addressed sign compression for DCT coefficients, a problem which was considered intractable in the image coding community. Inspired by the framework of phase retrieval~\cite{Bahmani2017flexible,Goldstein2018phasemax,Salehi18learning}, we formulated the problem of sign retrieval as a restoration of the underlying image from the magnitudes of the DCT coefficients. We developed a solution for this problem that alternates between DNN inference and POCS. The restored signs can be used to compress the sign information; only the difference between the restored and correct signs is sufficient to obtain the correct signs at the decoder side. The effectiveness of our method was validated through comparisons with previous methods in terms of the compression efficiency and computation cost.

Currently, our method is designed for still image coding and embedded into the pipeline of the standard JPEG. We believe that it can naturally be extended to video coding, where the transformation is performed on the residual images obtained through spatial and temporal predictions. We also need to consider other transformations than DCT, which are not always orthonormal. We believe this line of research can potentially boost the compression efficiency of video coding. We also expect that our interdisciplinary approach (a combination of phase retrieval and image coding) will inspire the community to further explore similar approaches, leading to new methodologies for image and video coding.

\bibliographystyle{IEEEtran}
\bibliography{main}

\end{document}